\documentclass[useAMS,usenatbib,asm]{mn2e}
\usepackage{graphicx,fleqn,times}

\usepackage{graphicx}
\usepackage{color}

\def\today{\ifcase\month\or
 January\or February\or March\or April\or May\or June\or
 July\or August\or September\or October\or November\or
 December\fi\space\number\day, \number\year}

\def\todmy{\number\day\space\ifcase\month\or
 January\or February\or March\or April\or May\or June\or
 July\or August\or September\or October\or November\or
 December\fi\space\number\year}

\title{Disc instabilities and semi-analytic modelling of galaxy formation}
\author[E. Athanassoula]
       {
       E. Athanassoula \\
Laboratoire d'Astrophysique de Marseille, Observatoire Astronomique de
  Marseille Provence,\\Technopole de l'Etoile - Site de Chateau-Gombert,
  38 rue Fr\'ed\'eric Joliot-Curie, 13388 Marseille C\'edex 13, France \\
}

\date{Accepted .
      Received ;
      }

\pagerange{\pageref{firstpage}--\pageref{lastpage}}
\pubyear{2006}

\begin{document}

\maketitle

\label{firstpage} 
\begin{abstract}
The Efstathiou, Lake and Negroponte (1982) criterion can not 
distinguish bar stable from bar unstable discs and thus should not be
used in semi-analytic galaxy formation simulations. I discuss the
reasons for this, illustrate it with examples and point out
shortcomings in the recipes used for spheroid formation. I propose an
alternative, although much less straightforward, possibility.
\end{abstract}


\section{Introduction}
\indent

Technically, it is not yet possible to make full cosmological
simulations which include both the dark matter and the baryons, 
down to the scale of individual galaxies. Yet
the dark matter only simulations have reached very high resolutions
(Springel et al. 2008, priv. comm.) and it is crucial to find ways of
exploiting 
them fully. A scheme involving resampling has been devised, in which a
given object and its surrounding region are singled out and
resimulated at high resolution (e.g. Katz \& White
1993). Nevertheless, this technique can only 
give information on one, or at best a few objects. For this reason,
semi-analytical models were introduced, which can give information on
properties of galaxy populations. These models are tagged on to 
the dark matter
only simulations and use `recipes' to describe the evolution of the
baryons. It is clear that the results will be useful only if the
recipes in question are correct and adequately chosen. This is not an
easy task, since these recipes need to include in a relatively simple
way a fair fraction of the available information on many key astrophysical
processes.  

In their quest to have more spheroids, either bulges or ellipticals,
some semi-analytic modellers are now considering  disc instabilities. 
They use the
Efstathiou, Lake and Negroponte (1982, hereafter ELN)) criterion to
distinguish between bar stable and bar unstable discs. Using
this criterion, this information can be obtained very simply from the
maximum rotational velocity and the disc mass and 
scale-length. Once a disc is found to be bar unstable, it is
turned instantaneously in the model into an elliptical (e.g. Bower et
al. 2006), or, alternatively, half of its mass 
is turned into a bulge and this repeatedly until disc stability
according to the ELN criterion is achieved (e.g. De Lucia \&
  Helmi 2008). In this way a
considerable amount of mass is turned into a spheroid and several
problems concerning the K-band luminosity function are 
alleviated (see Parry et al. 2008, for a discussion of the relative
merit of the two approaches). There are, however, a number of
shortcomings in this
modelling approach.

\section{Bar formation}
\indent

\begin{figure*}
  \setlength{\unitlength}{1cm}
  \includegraphics[scale=0.9,angle=0]{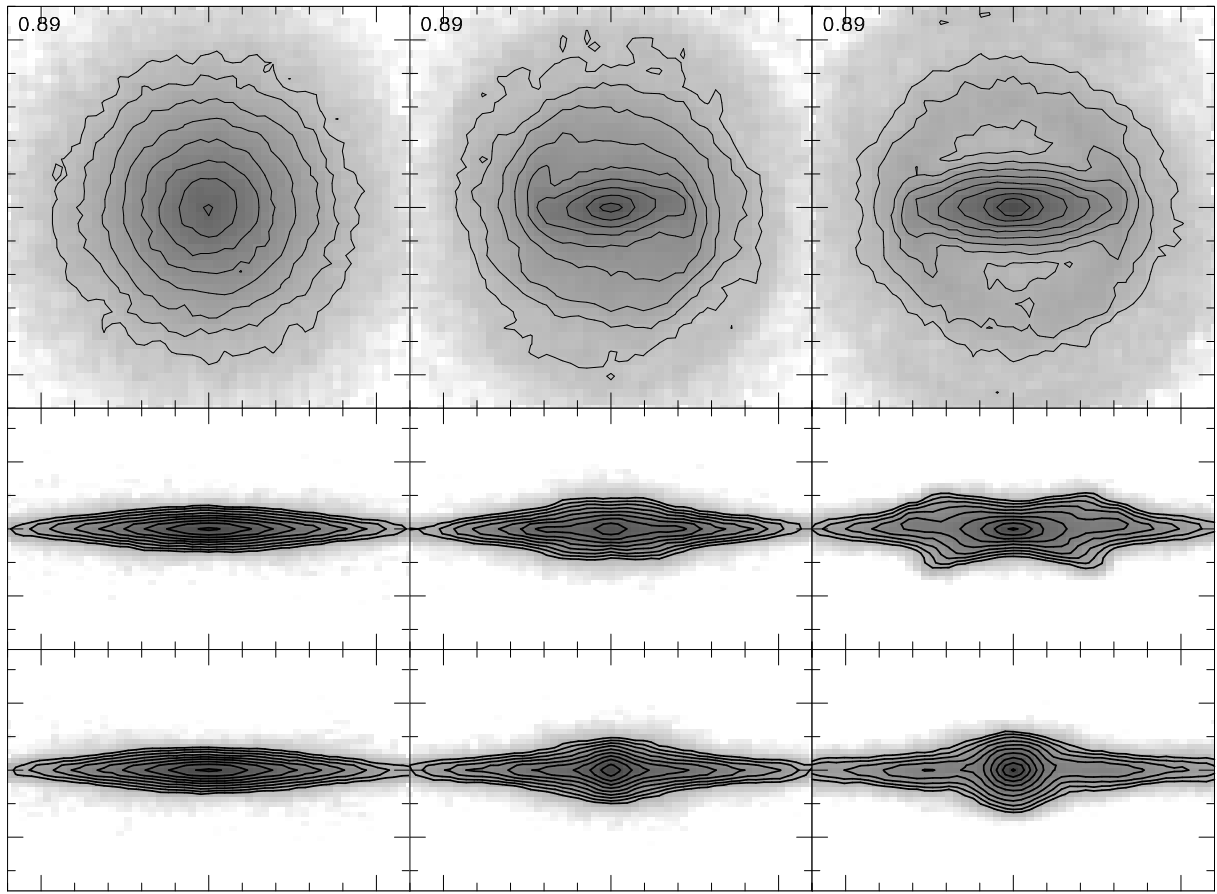}
\caption{The disc component of three simulations which, by the ELN
  criterion, should form a bar. The upper panels show the face-on
  views the middle ones the edge-on, side-on view (i.e. with the line
  of sight along the bar minor axis) and the lower ones the edge-on,
  end-on view (i.e. with the line of sight along the bar major
  axis). In all cases the projected density of the
  disc is given by
  grey-scale and also by isocontours (spaced logarithmically) and the
  numerical value of $R_{ELN}$ is given in the upper left corner
  of the face-on views.
}
\label{fig:unstable}
\end{figure*}

\begin{figure*}
  \setlength{\unitlength}{1cm}
  \includegraphics[scale=0.9,angle=0]{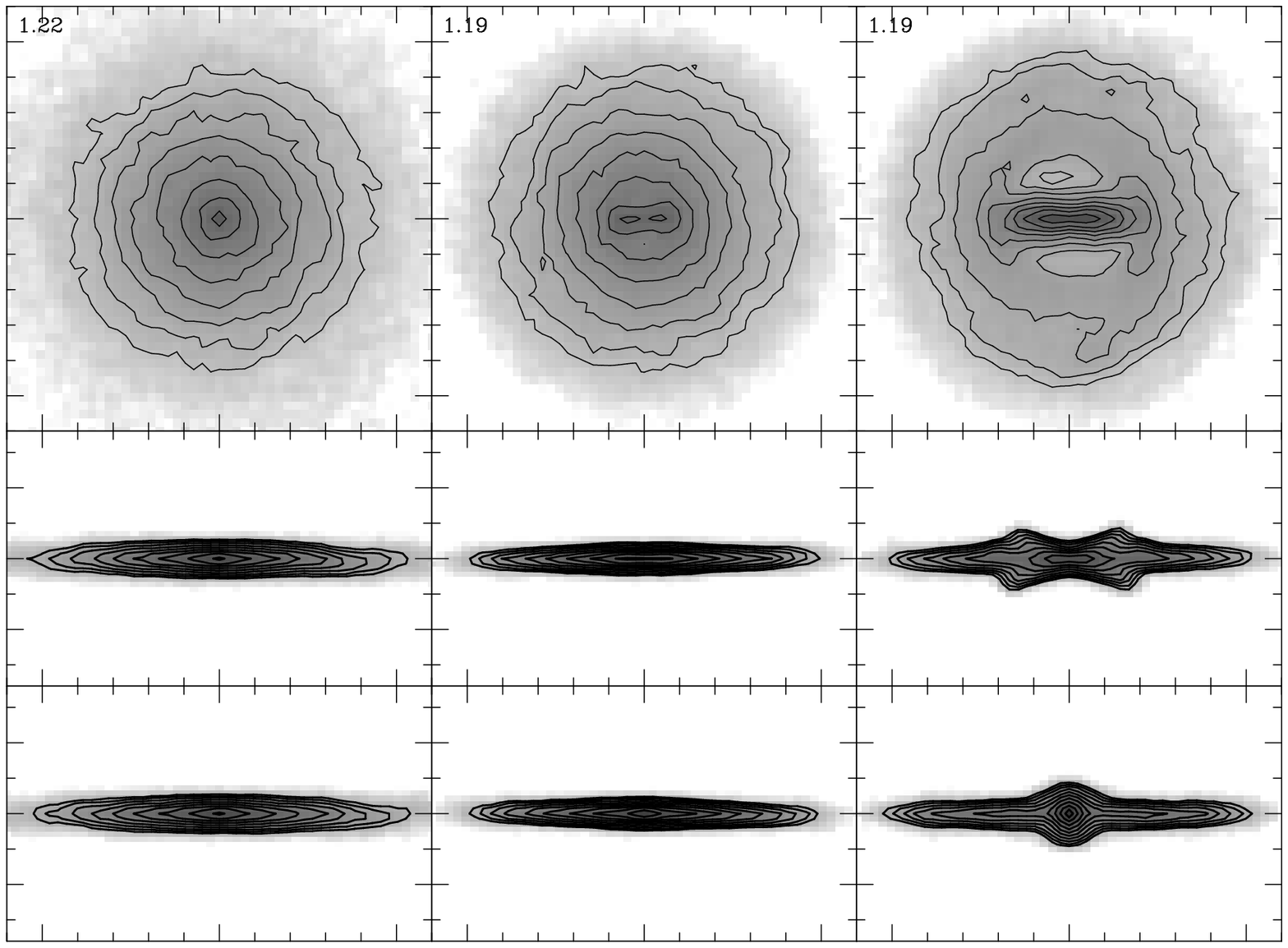}
\caption{Same as for Fig.~\ref{fig:stable}, but for three simulations
  which, by the ELN criterion, should not form a bar.
}
\label{fig:stable}
\end{figure*}

Efstathiou, Lake and Negroponte (1982) ran a number of two-dimensional
$N$-body
simulations of a purely stellar disc in a rigid halo and, based on
these, proposed a very simple criterion to distinguish bar stable from
bar unstable discs. This reads 
$$ R_{ELN} = {V_M} /{(M_D~G/R_D)^{1/2}},$$
\noindent
where $V_m$ is the maximum rotational velocity, $M_D$ is the mass of
the disc, $R_D$ is its scale length and $G$ is the gravitational
constant. If $R_{ELN} < 1.1$, ELN propose that the disc is bar
unstable, while it will be bar stable for values larger than that limit. 
This criterion was subsequently used by Mo, Mao \& White (1998) and is
sometimes referred to as the Mo, Mao \& White criterion. 

The ELN criterion was derived for purely stellar discs. Christodoulou,
Shlosman \& Tohline 
(1995) rederived it for the case of purely gaseous discs, and found
that the stability threshold is considerably lower. Thus, purely 
gaseous discs
will be stable if $R_{ELN} > 0.9$. So far, however, this criterion has
not been extended to the physically relevant case where both stars and
gas are available in the same disc, even when no star
formation or feedback is present. Our discussion will therefore, by
necessity, follow this limitation.  

Other variants of this criterion have also been proposed (see
e.g. van den Bosch 1998). In all its forms, however, the ELN criterion
is intimately related to the question of whether disks are self
gravitating in their inner parts, the so-called ``maximum disc'
problem (see Bosma 2004 for a review). 

The ELN criterion was derived more than 25 years ago and since then
our understanding of bar formation has advanced considerably. A number
of criticisms of this criterion can be and have been made. First the central
concentration of the halo has not been fully taken into account. In the linear
theory, if this was sufficiently high it would cut the swing amplifier
cycle and thus stop any growth of the bar mode (Toomre 1981). In
simulations, however, there can still be a tunnelling through this
barrier (e.g. Sellwood 1989), so that a bar may still grow. Thus this
criticism may not be particularly acute. 

A more serious one is that the disc velocity dispersion is not 
taken into account. A few years after the ELN criterion was published, 
Athanassoula \& Sellwood (1986) showed that velocity dispersion
has an important influence on bar stability, which is, in
fact, a function of both disc random motions and halo mass. They also
presented a disc with no halo, which is stable because of its high
velocity dispersion. Although these results are based on 2D
simulations, i.e. have by necessity rigid haloes, they still have the
advantage of stressing the effect of random motions on stability and
thus underline an inadequacy of the ELN criterion. 

A major problem of the ELN simulations is that they are two-dimensional. 
Such simulations have by definition rigid
haloes, i.e. haloes which are represented by an external forcing and
can not respond to the evolution of the disc. This approximation was
reasonable at the time, due to the limited computer means then
available. With the increase of
computer power, however, high resolution 3D simulations with adequate
self-consistent treatment of both the baryonic and the dark matter
component have become the norm. Athanassoula (2002) compared the results of
such fully self-consistent 3D simulations to those of simulations with
rigid haloes and showed that the results obtained with the latter
can be totally unreliable. In particular, she showed that rigid haloes
can erroneously claim stability for cases which the live haloes show 
are unstable. She explained the difference by the fact that angular momentum
exchange between the halo and the disc is {\it de facto} not existent
in simulations with rigid haloes. On the contrary, in fully
self-consistent simulations angular momentum is
emitted by near-resonant material in the bar region and absorbed by
near-resonant material in the halo and in the outer disc (Athanassoula
2003). This leads to a considerable growth of the bar and explains why
strong bars can be seen in discs immersed in massive haloes
(Athanassoula \& Misiriotis 2002). All this
complexity is of course not contained in the very simple ELN
criterion or in the work by Ostriker \& Peebles (1973), which come to the
opposite conclusions, since they do not include the 
interaction between the disc and the halo.

To make the inadequacies of the ELN criterion clearer, I have chosen
amongst my $N$-body simulations 
(Athanassoula 2003, 2007) six examples, three with $R_{ELN}$ smaller
than 1.1 and three with larger. The three first examples, shown in
Fig.~\ref{fig:unstable}, have identical $R_{ELN}$ values but different
disc velocity dispersion. Since
$R_{ELN}$ = 0.89, they should all three be bar unstable by the ELN
criterion, but Fig.~\ref{fig:unstable} shows that this is not the
case. The example on the left panel has a very hot disc which
makes it stable against bar formation, the middle one shows an
average-sized bar, while the one in the right panels shows a strong
bar. The value of $R_{ELN}$ chosen is nearly half way between the
minimum allowed value (0.63, for a bare exponential disc with no halo)
and the stability threshold of 1.1 and shows that even for 
$R_{ELN}$ values far from the stability threshold 
the disc velocity dispersion can stabilise the disc for
very long times, of the order of, or more than a Hubble time. This is
true for even lower values of $R_{ELN}$. For example, as already
stated, Athanassoula \& Sellwood (1986) produced a model with no halo
at all which was stable over a Hubble time. This is further
enhanced in the case where strong central concentrations, and/or high
velocity dispersions in the halo component add their stabilising
effect to that of the disc velocity dispersion. Thus, a galaxy may be bar
stable when the ELN criterion predicts instability.

The three examples given in Fig.~\ref{fig:stable} have
values of $R_{ELN}$ equal to 1.22 for the simulation on the left
panel, and 1.19 for the two others. Thus, the ELN criterion 
predicts that all three are bar stable, since their $R_{ELN}$ values
are larger than 1.1.  
Fig.~\ref{fig:stable} shows that one of the three cases is indeed bar
stable, the second one forms a small inner bar and the third one a
fair sized bar. The edge-on views are also widely different, since the
first two have only a disc component, while the third one shows a clear
peanut bulge. So again the ELN criterion is found faulty. This is due to
the fact that the live halo helps bar growth by absorbing at its
resonances the angular momentum that the inner disc can emit. Of
course, there is a limit beyond which the disc would not be able to
emit sufficient angular momentum, e.g. because it is not sufficiently
massive and/or because it is too hot. In such cases, the angular
momentum exchange within the galaxy would be limited by the emitters
(or lack thereof) and no bar could grow within an
astronomically relevant time even though the halo has responsive
resonances (Athanassoula 2003). This limit, however, is very far from
the ELN predictions, 
and, more important, does not depend only on the mass ratios, but also
on the velocity dispersions of the various components.

To summarise, the ELN criterion should not be used in semi-analytic
simulations. In cases where it predicts instability, the disc can
still be stabilised by 
factors which $R_{ELN}$ does not take into account, such as a strong central
concentration, strong random motions in the disc and halo, or, even
better, a combination of all three. Conversely, in cases where the ELN
criterion predicts 
stability, a bar can still form due to the destabilising influence of the
halo resonances. Shifting the threshold up or down will not solve the
problem as long all the other stabilising/destabilising influences are not
taken into account. Finally, note that in all cases, both bar stable
and bar unstable, the disc is preserved, i.e. no elliptical galaxy is
formed. 

\section{Summary and discussion}
\label{sec:summary}
\indent

In the previous section I showed 
that the ELN criterion is too simplistic to be
able to describe a complex phenomenon such as bar formation and to
distinguish bar stable from bar unstable discs. It thus cannot be
used in semi-analytic calculations. 

Further problems concern the subsequent evolution, after the bar has
formed. No simulation has ever shown an unstable disc turn into an 
elliptical, or acquire a very massive
classical bulge instantaneously. The bar can form a
boxy/peanut bulge (see Figs~\ref{fig:unstable} and \ref{fig:stable})
or a discy bulge, and the formation of a classical bulge from either
of those is not excluded\footnote{See Athanassoula (2005) for the
definition and properties of the different types of bulges.}. This 
classical bulge, however, would be much smaller than required by the
semi-analytic models, and, furthermore, it would not form
instantaneously since the bar needs some time to form, more time
is necessary for the boxy/peanut, or discy bulge formation and yet more
time is necessary for the putative conversion into a classical
bulge. Thus the second part of the De Lucia \& Helmi (2008) model has
shortcomings, but these may turn out to be quantitative rather than
qualitative. On the contrary, the Bower et al. (2006) model has
shortcomings at the qualitative level, since the disc can not 
disappear, i.e. an elliptical could not form from a bar unstable disc
without a merger. Both models of course have already
problems in the first part, i.e. deciding whether a disc is bar
unstable or not, since they both use the ELN criterion. 

Can we replace the simple ELN recipe for handling disc
instabilities by a better one? 
This task is not easy, since we need to include in this recipe
information on disc stability, bar formation and evolution and the
subsequent formation of the different types of bulges, all of which
are complex, 
nonlinear processes. I firmly believe that it is not possible to find
one, or a few simple formulas, like the ELN criterion, that can describe
all the necessary ingredients of these processes. It might thus be
preferable to seek a solution intermediate between the full
cosmological simulation including both the dark matter and the baryons
at sufficient resolution, which will not be available in the
foreseeable future, and the equally difficult task of finding
appropriate recipes. 

Computer simulations allow us today to routinely make $N$-body
simulations which can describe the evolution of a single galaxy.
In fact a very large number are already
available. Using their results instead of the recipes is a possibility
well worth exploring. For example, for the problem at hand one would
need a small library of $N$-body simulations with and without gas,
describing bar formation and evolution. These simulations should
cover, albeit very crudely, the necessary parameter space describing
the halo, stellar disc and gas components. This would include not only
the mass and scale-length ratios of the various components, but also
the different amounts of random motion in the stellar disc and in the
spheroids. From these 
simulations one should extract all the necessary parameters, such as
the times necessary for bar and peanut formation, the amount of mass
in the boxy/peanut bulge or the discy bulge, etc. This
information would be assembled in some sort of table. Then at 
each time step of the semi-analytic simulation, instead of checking the 
criterion or applying the recipe, one would extract from the above
table the relevant information as a function of the properties of the
galaxy at the time under consideration. 

The above scheme is of course not a substitute for full scale cosmological
simulations, starting {\it ab initio}. It also has a number of
shortcomings, the most important of which is that our knowledge about
the velocity dispersions in disc galaxies is very limited. 
Nevertheless, it is a very important improvement with respect to
the presently used scheme.

\section*{Acknowledgements}

I thank Albert Bosma and Gabriella De Lucia for useful discussions and Simon
White and Albert Bosma for comments on the manuscript. I also thank
Guinevere Kauffmann and Dimitri Gadotti for inviting me to MPA where this work
started. This work was partially supported by grant ANR-06-BLAN-0172. 

\bibliographystyle{mn2e}

\bibliography{semianalELN_ansref}

\label{lastpage}

\end{document}